\documentclass[%
 aip,
 amsmath,amssymb,
 reprint,%
]{revtex4-1}

\usepackage{graphicx}
\usepackage{dcolumn}
\usepackage{bm}

\usepackage[utf8]{inputenc}
\usepackage[T1]{fontenc}
\usepackage{mathptmx}
\usepackage{etoolbox}

\makeatletter
\def\@email#1#2{%
 \endgroup
 \patchcmd{\titleblock@produce}
  {\frontmatter@RRAPformat}
  {\frontmatter@RRAPformat{\produce@RRAP{*#1\href{mailto:#2}{#2}}}\frontmatter@RRAPformat}
  {}{}
}%
\makeatother
\begin{document}

\preprint{AIP/123-QED}

\title{Optimizing GaAs/AlGaAs Growth on GaAs (111)B for Enhanced Nonlinear Efficiency in Quantum Optical Metasurfaces}



\author{T. Blaikie}
\affiliation{ 
Electrical and Computer Engineering, University of Waterloo, 200 University Ave. W, Waterloo, ON N2L 3G1, Canada
}%

\author{M. C. Tam}
\affiliation{ 
Electrical and Computer Engineering, University of Waterloo, 200 University Ave. W, Waterloo, ON N2L 3G1, Canada
}%

\author{S. Stich}
\affiliation{ 
Walter Schottky Institute Technical University of Munich, Am Coulombwall 4, 85748 Garching bei München, Germany
}%

\author{M. Belkin}
\affiliation{ 
Walter Schottky Institute Technical University of Munich, Am Coulombwall 4, 85748 Garching bei München, Germany
}%

\author{Z. R. Wasilewski}
\affiliation{ 
Electrical and Computer Engineering, University of Waterloo, 200 University Ave. W, Waterloo, ON N2L 3G1, Canada
}%
\affiliation{ 
Physics and Astronomy, University of Waterloo, 200 University Ave. W, Waterloo, ON N2L 3G1, Canada
}%
\affiliation{Waterloo Institute for Nanotechnology, 200 University Ave. W, Waterloo, ON N2L 3G1, Canada}

\date{\today}

\begin{abstract}
This study is an optimization of GaAs and Al$_{0.55}$Ga$_{0.45}$As growth on GaAs (111)B substrates with a surface misorientation of 2° towards [$\bar{2}$11], aiming to enhance surface morphology. For quantum optical metasurfaces (QOMs) the change from (001) growth to (111) growth will increase the efficiency of the nonlinear process of spontaneous parametric downconversion (SPDC). This work identifies key factors affecting surface roughness, including the detrimental effects of traditional thermal oxide desorption, which prevented growth of smooth surfaces. A novel Ga-assisted oxide desorption method using Ga flux ``pulses'' was developed, successfully removing oxides while preserving surface quality. In-situ characterization techniques, including reflection high energy electron diffraction (RHEED) and a novel technique called Diffuse Laser Scattering (DLS), were employed to monitor and control the oxide removal process. Mounting quarter wafers with sapphire substrates as optical diffusers improved surface uniformity by mitigating temperature gradients. The uniformity of the growths was clearly visualized by an in-house technique called black box scattering (BBS) imaging.  Indium (In) was tested as a surfactant to promote step-flow growth in GaAs, which resulted in atomic-scale flatness with root mean square (RMS) roughness as low as 0.256 nm. In combination with a higher growth temperature, the RMS roughness of Al$_{0.55}$Ga$_{0.45}$As layers was reduced to 0.302 nm. The optimized growth methods achieved acceptable roughness levels for QOM fabrication, with ongoing efforts to apply these techniques to various devices based on GaAs/AlGaAs heterostructures with (111) orientation.
\end{abstract}

\maketitle

\section{\label{sec:Introduction} Introduction}

Nonlinear optical phenomena such as second-harmonic generation (SHG) and spontaneous parametric down-conversion (SPDC) arise from interactions between light and dielectric materials with nonlinear electric susceptibilities. Nonlinear optical devices utilize SPDC for photonic quantum state engineering but are typically limited by strict momentum conservation. Instead, nonlinear materials can be fabricated into quantum optical metasurfaces (QOMs), thin layers patterned with arrays of nanoresonators specifically tuned for optical resonances. The subwavelength thickness of QOMs relaxes the constraint of momentum conservation between the photons involved in SPDC, allowing for greater versatility of the produced quantum states.

 SHG and SPDC are second-order nonlinear optical processes, therefore the emission rates of these processes depend on the second-order electrical susceptibility ($\chi^{(2)}$). Gallium arsenide (GaAs) has one of the highest second-order electrical susceptibilities of traditional materials, which promotes its use in nonlinear optics and quantum photonics. 
 
 
 In a recent study, metasurfaces were fabricated from thin layers of GaAs that were grown by molecular beam epitaxy (MBE) on GaAs (001) substrates\cite{santiago-cruz_resonant_2022}. Strong quasi-bound in continuum (qBIC) resonances were identified with high quality (Q) factors of $\approx$330 and $\approx$1000 for electric dipole and magnetic dipole modes, respectively. The efficiency of entangled photon pair emission was at least three orders of magnitude greater when compared to an unpatterned GaAs layer of the same thickness. These QOMs pave the way for advancements in nanoscale photonic quantum state engineering and the miniaturization of photonic quantum processing.

 

In general, the susceptibilities of dielectric materials can be anisotropic, leading to different effects depending on the relative orientation between the applied electric field and the dielectric material. The susceptibility of GaAs is anisotropic and is represented by the $\chi^{(2)}$ tensor. For GaAs (111), the elements of $\chi^{(2)}$ are more favourably aligned than for (001), enhancing coupling to light that is orthogonal to the surface. 
 
 Metasurfaces fabricated on GaAs (001) require designs that include at least one resonant mode capable of generating a significant vertical electric field component. Achieving this is challenging, and the resulting nonlinear downconversion efficiency is relatively low. Conversely, metasurfaces fabricated on GaAs (111) can deliver a strong nonlinear response when the electric fields of all three interacting photons lie within the plane of the metasurface. 
 
 Simulations of the same GaAs (001) structures as in Ref. [\onlinecite{santiago-cruz_resonant_2022}], except for with GaAs (111), show that SPDC efficiency could increase by 1 to 3 orders of magnitude for wavelengths of interest. Note that these designs are not optimized for GaAs (111) and further enhancement would be expected with new designs intended for GaAs (111)B.

In this work, we focus on optimizing the growth of GaAs/AlGaAs heterostructures on GaAs (111)B substrates.

Notably, most studies on GaAs (111) growth were published prior to 2000. Achieving high-quality (111) growth requires precise control of growth conditions to suppress hillock formation. This challenge is further complicated by the surface reconstructions of the (111) plane, which are less understood than those observed in (001) growth. Due to these difficulties, growth along the (111) direction has historically been considered very challenging, leading to a shift in focus toward the (001) orientation. In recent years, there has been renewed interest in (111) growths because new technologies, such as QOMs, could potentially benefit from the change in crystal orientation. 

The objective of our optimization was to produce materials with minimal surface roughness and minimal defects. These qualities are essential for reducing scattering effects in optical experiments and improving metasurface fabrication. In the literature, there is no consensus on the optimal growth conditions to achieve a smooth surface morphology when growing on GaAs (111)B substrates. Nevertheless, several papers on the growth of GaAs/AlGaAs heterostructures on GaAs (111) substrates were examined to establish a starting point for our optimization efforts\cite{shitara_rheed_1990,hayakawa_reduction_1987,hayakawa_molecular_1991,takano_growth_1990,tsutsui_optimum_1990,chen_growth_1991,chen_relation_1991,chin_high_1991,yang_molecular-beam_1993,guerret-piecourt_temperature_1998,herzog_optimization_2012,grey_growth_1995}.


\section{\label{sec:Experiment}Experiment}
A series of epitaxial growths were performed to study the impact of various growth parameters and procedures on the quality of GaAs and Al$_x$Ga$_{1-x}$As layers. These growths were carried out using a Veeco Gen10 MBE system equipped for arsenide and antimonide growths. Several growth conditions were kept constant throughout the study. The As cracking cell was configured to produce an As$_4$ flux. The Group V/III flux ratio was maintained above 2, ensuring that the As overpressure was approximately double the minimum required to sustain an As-stable surface reconstruction. The GaAs growth rate ($R_{GaAs}$) was set at 1.0 Å/s.

Each growth was carried out on quarters of 3" wafers. The wafer pieces were held stationary during growth. This approach simplified in-situ monitoring and allowed for some variation of growth conditions across the wafer piece, enabling the evaluation of different V/III flux ratios. For our substrates, we chose epi-ready GaAs (111)B wafers with a nominal surface misorientation of 2° towards [$\bar{2}$11]. This intentional surface misorientation is a crucial parameter for suppressing hillock formation during growth.

The surface misorientation from the (111) plane creates a vicinial surface with atomically flat terraces separated by atomic-height steps. This facilitates step-flow growth, where adatoms preferentially settle at kink and step sites along the edges of terraces, as noted in Ref. [\onlinecite[p.~42]{ida_sadeghi_realization_2021}]. However, the surface misorientation of our substrates was measured to be slightly off-target at 2.4° in magnitude. 

The conditions for each growth are shown in Table \ref{Tab:Growths}. The quarter wafer from each growth is identified (ID) by ``q-'' followed by a capital letter from A-I (ex. q-A). Each growth from q-A to q-F underwent oxide desorption followed by the growth of 300 nm of GaAs. q-G to q-I had an additional 300 nm thick layer of Al$_x$Ga$_{1-x}$As grown on top of the GaAs layer.


\begin{table*}
    \caption{\label{Tab:Growths} This table shows the growth conditions for the series of growths in the optimization study. For all of the growths, the surface misorientation is 2° toward [$\bar{2}$11], and the As$_4$ flux is set high enough that the group V/III flux ratio is always greater than 2. $R_{GaAs}$, $R_{Al_{1-x}Ga_xAs}$ and $R_{InAs}$ are the nominal growths rates used for GaAs, Al$_{1-x}$Ga$_x$As, and InAs respectively. $N_{Ga}$ is the number of Ga pulses used during oxide desorption. $T_G$ is the growth temperature.  $R_{RMS}$ is the RMS roughness measured by an AFM scan, of size S, in the smoothest region of the wafer piece. Usually the smoothest region is the center, see FIG. \ref{fig:Progression}.}
    \centering
    \begin{ruledtabular}
    \begin{tabular}{cccccccccc}
    ID & Technique                                                                               & \begin{tabular}[c]{@{}c@{}}Oxide Desorption\\ Method\end{tabular} & $N_{Ga}$ & $T_G$ (°C)      & \begin{tabular}[c]{@{}c@{}}$R_{GaAs}$\\ (Å/s)\end{tabular} & \begin{tabular}[c]{@{}c@{}}$R_{Al_{1-x}Ga_xAs}$\\ (Å/s)\end{tabular} & \begin{tabular}[c]{@{}c@{}}$R_{InAs}$\\ (Å/s)\end{tabular} & \begin{tabular}[c]{@{}c@{}}$R_{RMS}$\\ (nm)\end{tabular} & \begin{tabular}[c]{@{}c@{}}S\\ (µm$^2$)\end{tabular} \\
    \hline
    q-A  & High Temperature Oxide Desorption & Thermal                                                           & 0         & 630       & 1.0                                                         & --                                                            & --                                                           & 3.26                                                      & 10                                                                           \\[0.32cm]
    q-B  & Low Temperature Oxide Desorption   & Thermal                                                           & 0         & 590       & 1.0                                                         & --                                                            & --                                                          & 1.40                                                      & 1                                                                            \\[0.32cm]
    q-C  & Ga-Assisted Oxide Desorption       & Ga-Assisted                                                       & 7         & 590       & 1.0                                                         & --                                                            & --                                                          & 0.697                                                     & 1                                                                            \\[0.32cm]
    q-D  & Mounted with a Sapphire Diffuser  & Ga-Assisted                                                       & 7         & 590       & 1.0                                                         & --                                                            & --                                                          & 0.917                                                     & 1                                                                            \\[0.32cm]
    q-E  & High Flux Indium Surfactant        & Ga-Assisted                                                       & 6         & 590        & 1.0                                                         & --                                                            & 1.15                                                        & 2.50                                                      & 1                                                                            \\[0.32cm]
    q-F  & Low Flux Indium Surfactant         & Ga-Assisted                                                       & 6         & 590        & 1.0                                                         & --                                                            & 0.20                                                        & 0.256                                                     & 1                                                                            \\[0.32cm]
    q-G  & Low Temperature AlGaAs Layer            & Ga-Assisted                                                       & 6         & 590        & 1.0                                                         & 2.23                                                          & 0.20                                                        & 0.548                                                     & 1                                                                            \\[0.32cm]
    q-H  & High Temperature AlGaAs Layer     & Ga-Assisted                                                       & 6         & 590/630       & 1.0                                                         & 2.23                                                          & 0.20                                                         &  0.302                                                     & 1                                                                           \\[0.32cm]
    q-I  & No In Flux During AlGaAs Layer     & Ga-Assisted                                                       & 6         & 590       & 1.0                                                         & 2.23                                                          & 0.20                                                         &  0.599                                                     & 1                                                                           
    \end{tabular}
    \end{ruledtabular}    
\end{table*}

\subsection{In-Situ Monitoring}

Reflection high energy electron diffraction (RHEED) was used to observe surface reconstruction during each growth stage. Also, a RHEED signal was monitored in real-time by integrating the intensity of the pixels on the specular reflection spot. All substrate temperatures reported in this work were measured using band-edge thermometry (BET)\cite{johnson_semiconductor_1993,johnson_real-time_2000}. Group III fluxes were measured prior to growth with a beam flux monitoring (BFM) ion gauge positioned directly beneath the nominal substrate location.

\begin{figure}
\centering
\includegraphics[width=\columnwidth]{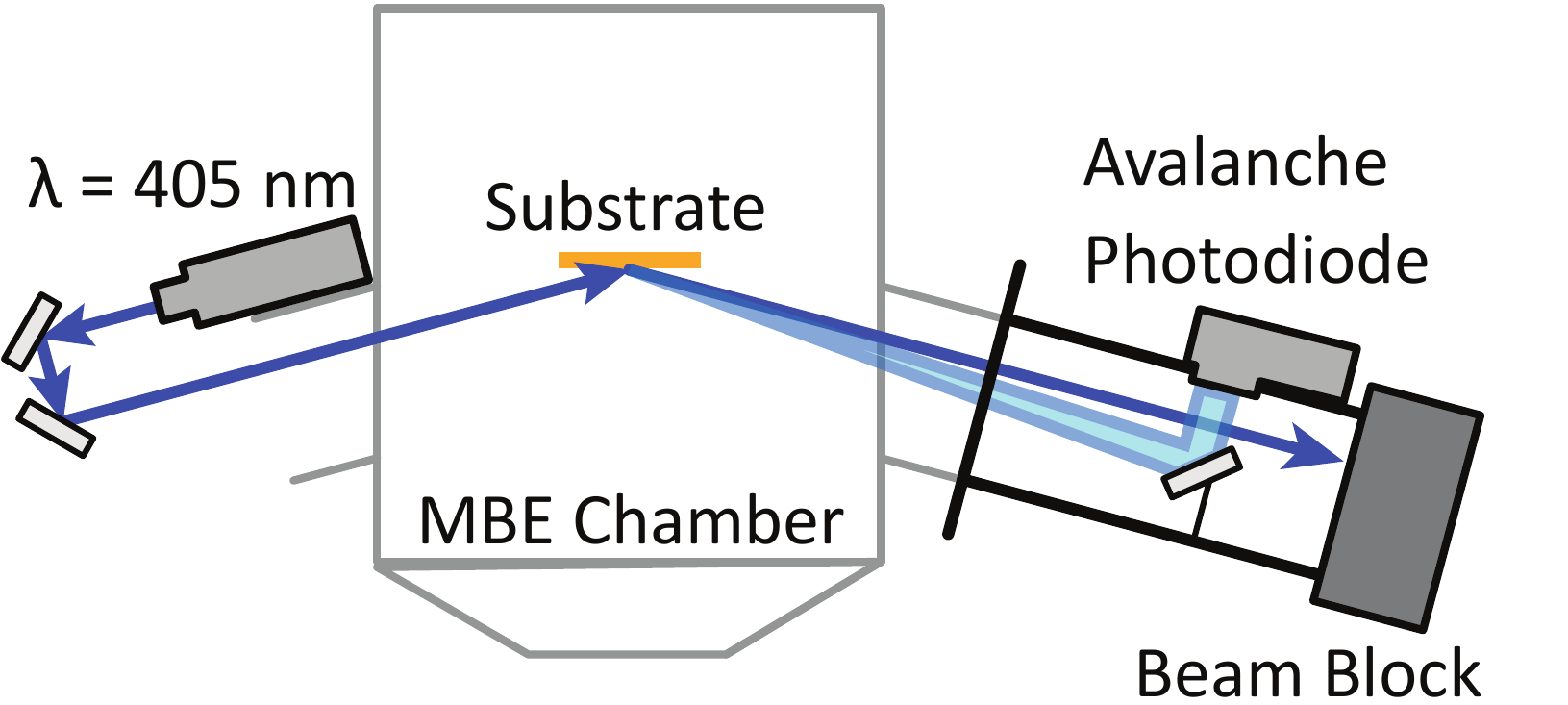}
\caption{A schematic of the optical setup for measuring diffuse laser scatter signal. This configuration measures the intensity of light that has been scattered with a very small deviation from the specular beam.}
\label{fig:DLSSetup}
\end{figure}

Additionally, an in-house tool was developed to measure the diffuse scattering of light from the growth surface. This is referred to as diffuse laser scatter (DLS) measurement. The light source is a laser diode emitting at 405 nm wavelength. FIG. \ref{fig:DLSSetup} presents a schematic of the optical setup for this measurement. Two viewports on opposite sides of the MBE growth chamber allow light to enter and exit the chamber. On the first viewport, the laser is mounted along with two mirrors, allowing the laser beam to be directed through the window. The beam is incident at the center of the substrate with a shallow incident angle of $\sim$15°. The specular beam exits the chamber through the opposite viewport and is stopped by a beam-absorbing block. The specular beam is not monitored; instead, the measured signal comes from diffusely scattered light.

There are two configurations of this setup to collect the diffusely scattered light. The primary configuration, used for most of the data in this report, relies on a mirror placed just outside of the specular beam path at an angle close to 45°, to redirect diffusely scattered light into an avalanche photodiode (APD). This is pictured in FIG. \ref{fig:DLSSetup}. The APD signal is then measured using a Stanford Research Systems SR850 lock-in amplifier. A 100 kHz reference signal was used to modulate the laser power digitally. This configuration measures light intensity scattered at angles close to the specular beam. The second possible configuration uses a viewport on the bottom of the growth chamber. Due to limited space around this viewport, a fiber optic cable is used to collect the light, which is then transmitted to the same APD and lock-in amplifier. This configuration measures light scattered at a large angle, far from the specular beam. The DLS signal can be measured while the substrate is rotating, but any substrate wobble present during rotation will complicate the analysis. For this study, the wafers were kept stationary.

This tool has been implemented primarily to evaluate the evolution of surface roughness during growth with high sensitivity. The intensity of diffusely scattered light increases with surface roughness. This is the main principle with which the DLS measurements have been qualitatively evaluated. However, the signal intensity is also influenced by factors such as the geometry of the optical setup, the polarization of the light, and the temperature-dependent dielectric constants of the materials. We did not investigate these effects; instead, we monitored the signal for changes in intensity that corresponded to changes in growth conditions.

\subsection{Ex-Situ Characterization}

Nomarski differential interference contrast (DIC) microscopy was employed for rapid evaluation of surface morphology over a large area. Additionally, an atomic force microscope (AFM) was used to visualize the surface of each growth across areas as small as 1 µm$^2$ and to measure RMS roughness with sub-nanometer vertical resolution.

An additional in-house characterization technique, called black box scattering (BBS), was utilized to evaluate the uniformity of the surface roughness across the entire wafer for each growth. This simple technique involves placing the wafer on a stand inside a box with black light-absorbing walls on all sides. A powerful remote camera flash is mounted to the top of the box, pointing straight down at the wafer. A camera positioned at a 45° angle from the wafer surface captures an image synchronized with flash illumination.

If the wafer is defect-free and very smooth, the BBS image will show a completely black surface. In this case, the light from the flash is specularly reflected toward the top of the box and it is not captured by the camera. However, if the wafer has defects or some roughness, these will be revealed in the image as brighter areas because the light from the flash will be scattered towards the camera. We take these images with the wafer in multiple orientations, as the roughness is often anisotropic, and some directions will show more scattered light than others.

\begin{figure}
\centering
\includegraphics[width=\columnwidth]{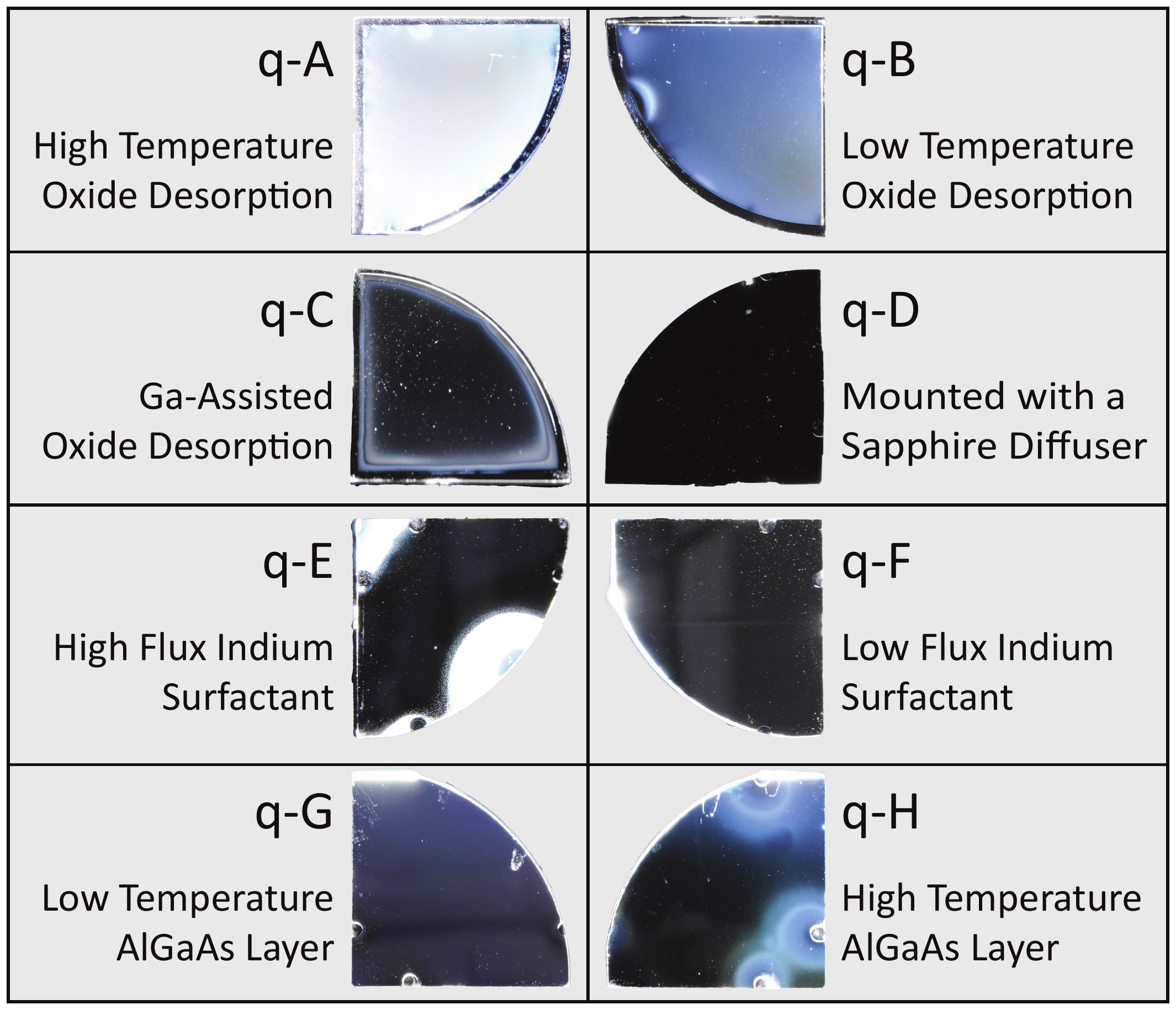}
\caption{Photos of quarter wafers corresponding to the growths listed in Table \ref{Tab:Growths}, captured using the BBS technique. Under normal inspection with the naked eye these wafers would appear to have smooth specular surfaces. The BBS imaging technique is a powerful way to reveal small scale roughness and defects across a wafer's surface. The photo for q-I is omitted because it looks nearly identical to q-G.}
\label{fig:Progression}
\end{figure}

For each growth in Table \ref{Tab:Growths}, a corresponding photo of the quarter wafer is shown in FIG. \ref{fig:Progression}, taken using the BBS technique. It is important to note that all of these wafers, except for the roughest one, q-A, would appear to have smooth specular surfaces to the naked eye. The haze seen in most BBS images is a testament to how powerful this technique is in revealing even the smallest defects and deviations from atomic smoothness.


Finally, X-ray diffraction (XRD) techniques were used to verify material composition and layer thickness for growths that included Al$_x$Ga$_{1-x}$As layers. The diffractometer used was a high-resolution Bruker QC3 system. Symmetrical omega-2theta scans around the 222 Bragg peak were measured, and the diffractograms were analyzed using Bruker's RADS software, which is based on the dynamical theory of X-ray diffraction.

\subsection{Growths}

Table \ref{tab:Schematic} provides an example structure that could be used to fabricate QOMs. The composition and thickness of the primary layer may be chosen differently for the final QOMs, but the buffer and etch stop layers are essential for transferring the primary layer to a transparent substrate for optical measurements\cite{santiago-cruz_resonant_2022}. Therefore, the main challenge is optimizing the growth of GaAs and of Al$_{0.55}$Ga$_{0.45}$As for the etch stop layers.

\begin{table}
\caption{An example structure for MBE growth for QOM fabrication.}
\label{tab:Schematic}
\begin{ruledtabular}
\begin{tabular}{ccc}
Description			& Material			            & Thickness (nm) 	    \\
\hline
Substrate			& SI-GaAs (111)B 		        & $6.25 \times 10^5$	\\
Buffer Layer		& GaAs				            & 200				    \\
Etch Stop 1		    & Al$_{0.55}$Ga$_{0.45}$As		& 300				    \\
Etch Stop 2		    & GaAs				            & 100    				\\
Primary Layer       & Al$_{0.30}$Ga$_{0.70}$As        & 1800                   \\

\end{tabular}
\end{ruledtabular}
\end{table}

\subsubsection{Thermal Oxide Desorption}

The first issue identified with our growth procedure was related to thermal oxide desorption. Our typical oxide desorption procedure for GaAs (001) substrates was to heat the wafer to 630 °C and anneal it for 20 minutes under As flux. Applying such a procedure to GaAs (111)B epi-ready substrates resulted in wafers with unsatisfactory surface morphology. The AFM scans showed substantial surface roughness, and the BBS image's haziness indicated that this roughness was present over the entire wafer. Several attempts were made with different As overpressures, growth temperatures, and growth rates, searching for optimized conditions. Each growth resulted in a surface that was either comparably rough to the initial growths or significantly worse (in the case of too low a V/III ratio). Even when the Al$_{0.55}$Ga$_{0.45}$As layers were eliminated, and only the GaAs buffer layer was grown, the surface was always rough. 

Monitoring RHEED during these growths revealed that the surface oxide starts to desorb at around 590 °C. For each GaAs (111)B growth, before heating the wafer, RHEED would begin with a diffuse, hazy pattern. As the substrate temperature increased, the diffused background would fade slightly, and at 590 °C, the background would drop significantly. Over the next 30 seconds, a pattern of spots and streaks emerged, indicating the desorption of surface oxide.

\begin{figure}
\centering
\includegraphics[width=\columnwidth]{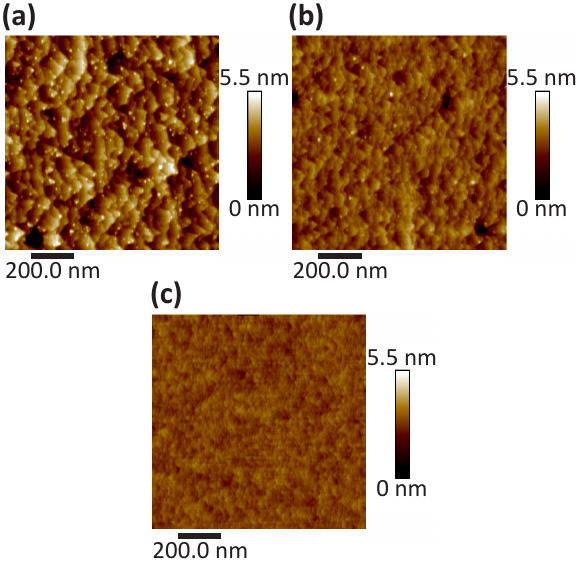}
\caption{(a-c) 1 µm$^2$ AFM scans from the center of quarter wafers. These scans show wafers that underwent oxide desorption without subsequent growth. (a) High-temperature (630 °C) thermal oxide desorption with R$_{RMS}$ = 0.770 nm. (b) Low-temperature (590 °C) thermal oxide desorption with R$_{RMS}$ = 0.436 nm. (c) Ga-assisted oxide desorption with R$_{RMS}$ = 0.230 nm.}
\label{fig:OxideDesorptionAFM}
\end{figure}

To investigate the effect of oxide desorption temperature on surface morphology, a GaAs (111)B wafer was cleaved into four quarters, and two of the quarters were used. Each quarter wafer was heated up and annealed under As flux, with no subsequent deposition. Their surface morphologies were assessed using 1 µm$^2$ AFM scans (FIG. \ref{fig:OxideDesorptionAFM}). The quarter wafer in FIG. \ref{fig:OxideDesorptionAFM} (a) went through our standard oxide desorption procedure, being heated to 630°C and annealed for 20 minutes, resulting in RMS roughness of 0.770 nm. The quarter wafer in FIG. \ref{fig:OxideDesorptionAFM} (b), on the other hand, was heated only to 590°C and held at that temperature for 15 minutes, resulting in RMS roughness of 0.436 nm. This shows that the higher temperature thermal oxide desorption process of GaAs (111)B roughens the surface significantly more than the lower temperature oxide desorption.


A similar comparison between two substrates with subsequent growths of 300 nm of GaAs also showed lower RMS roughness for the lower temperature thermal oxide desorption. FIG. \ref{fig:Progression} shows BBS images for q-A and q-B. The RMS roughness ($R_{RMS}$) from AFM scans of the smoothest region of each quarter wafer are given in Table \ref{Tab:Growths}. The quarter q-A was subject to a standard high-temperature oxide desorption (630 °C for 20 minutes), and then 300 nm of GaAs was grown at 630 °C. The quarter q-B had a reduced oxide desorption annealing temperature (590 °C for 15 minutes), and then 300 nm GaAs layer was grown at 590 °C. It is evident from the BBS images that neither sample has a specular surface, but q-B is better than q-A. We attribute that to the growth on q-B starting on a smoother surface, with fewer deviations from the desired surface misorientation.

The surface roughening, in both cases, results from surface reactions taking place during thermal oxide desorption. For epi-ready substrates like the ones used in this study, the wafer manufacturer prepares the surface, and a long time passes before the wafer is used for growth. During that time, gallium oxide (Ga$_2$O$_3$) forms through a spatially inhomogeneous reaction between the arsenic oxide (As$_2$O$_3$) and the GaAs surface\cite{wasilewski_studies_2004}:

\begin{center}
2GaAs + As$_2$O$_3$ $\rightarrow$ 4As + Ga$_2$O$_3$.
\end{center}
During thermal oxide desorption, the thermally stable Ga$_2$O$_3$ is converted into volatile Ga$_2$O through another reaction with the surface:
\begin{center}
Ga$_2$O$_3$ + 4GaAs $\rightarrow$ 3Ga$_2$O + 2As$_2$ (or As$_4$).
\end{center}
This reaction consumes GaAs from the substrate surface, creating significant roughness and an inhomogeneous distribution of deep pits in the surface\cite{wasilewski_studies_2004}, like those seen in FIG. \ref{fig:OxideDesorptionAFM} (a).

This roughness is detrimental to (111) growth specifically, where optimized surface misorientation is essential for a step-flow growth mode. The pits in the surface expose many facets in directions different from the preferred surface orientation. Unlike in the case of epitaxy on (001) substrates, a smooth surface cannot be recovered by growing a buffer layer on the (111) plane because of the sensitivity of growth modes to surface misorientation. This explains why our initial attempts at epitaxial growth on GaAs (111)B surface were unsuccessful.

\subsubsection{Ga-Assisted Oxide Desorption}

\begin{figure*}
\centering
\includegraphics[width=\textwidth]{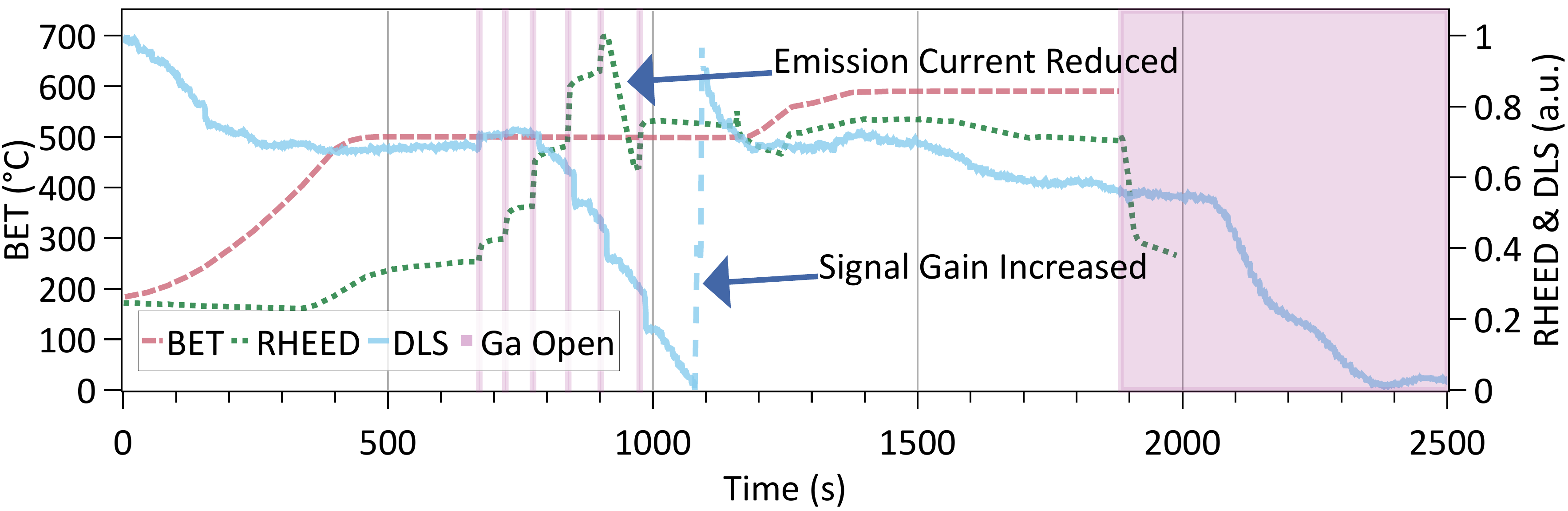}
\caption{Graphical summary of a representative Ga-assisted oxide desorption process followed by subsequent GaAs growth. The plot shows BET, RHEED and DLS signals. Shaded regions are where the Ga cell shutter was open.}
\label{fig:GraphSummary}
\end{figure*}

The oxide removal procedure was further improved by using a Ga-assisted oxide desorption technique, similar to the method proposed in Ref. [\onlinecite{wasilewski_studies_2004}]. The RHEED, BET, and DLS signals, as illustrated in FIG. \ref{fig:GraphSummary}, are presented for a representative oxide desorption process followed by subsequent growth of GaAs. 

The Ga-assisted oxide desorption procedure follows these steps: First, the substrate is ramped up to 500 °C without As flux, well below the temperature needed for rapid oxide conversion via reaction with the substrate. Several ``pulses'' of Ga are then sent to the wafer by opening the Ga cell shutter for 2 seconds. The Ga cell temperature is set for a growth rate of 1.0 Å/s, so 2 seconds deposits one monolayer of Ga. 

In contrast to the standard thermal oxide desorption, these pulses directly supply the Ga needed for the reaction rather than obtaining it from Ga-As bonds in the substrate, with the latter requiring extra energy and, thus, higher temperatures. These pulses are repeated with a 40-second interval between each. Depending on the specific wafer and the amount of oxide present, 6-8 pulses may be required to convert most of the oxide into volatile Ga$_2$O. 

It is important to avoid sending too many Ga pulses, as excessive pulses can result in Ga droplets on the surface. Therefore, it is best to remove \textit{most} of the oxide using Ga pulses and then remove the remaining oxide through the typical reactive oxide conversion by increasing the substrate temperature.

Our in-situ characterization tools are the key to identifying a safe number of pulses in each case. In FIG. \ref{fig:GraphSummary}, the shaded regions indicate the time when the Ga cell shutter was opened. A clear response to the Ga pulses is observed in both the RHEED and DLS signals between 600 and 1000 seconds. After five Ga pulses, the emission current of the electron gun had to be reduced to prevent camera saturation. 

In this example, the RHEED signal intensity increased for all six Ga pulses. The increasing intensity of the RHEED pattern's specular reflection spot is accompanied by the reduction of the diffuse background and the emergence of a streaky pattern. The DLS signal increased slightly for the first two Ga pulses but dramatically decreased for the final four pulses. The initial increase in the DLS signal suggests that the surface became rougher, possibly due to the formation of Ga droplets, but these were ultimately consumed in reaction with the oxides. 

The gain of the APD for the DLS signal had to be increased after the Ga pulses to continue monitoring surface morphology. Based on experience, one can judge when to stop sending Ga pulses. If even one too many Ga pulses are sent, it will manifest as a decrease in RHEED intensity and/or an increase in the DLS signal.

After the Ga pulses are completed, the substrate temperature is raised to 590°C for 10 minutes under As overpressure to remove residual oxides. Some of the RHEED and DLS signal fluctuations during this phase are attributed to surface reconstruction evolution and small wafer movements as the manipulator temperature is increased.

The deposition of GaAs starts at around 1900 seconds in FIG. \ref{fig:GraphSummary}. The RHEED signal drops as the surface reconstruction changes again. This particular growth involved co-deposition of In surfactant, as discussed later. The RHEED and BET viewport shutters were kept closed during most of the epitaxial growth phase to prevent the viewports from being coated with In re-evaporated from the surface. However, the DLS signal was monitored throughout the growth and continued its downward trend, indicating that the surface was becoming smoother.

To evaluate the effectiveness of the Ga-assisted oxide removal procedure, another quarter wafer was used with only the oxide removal procedure applied and no subsequent growth. As shown in FIG. \ref{fig:OxideDesorptionAFM} (c) the resulting surface after Ga-assisted oxide removal is very smooth with RMS roughness of just 0.230 nm. This provides a significantly better surface condition for GaAs deposition on (111)B substrate, compared to thermal oxide desorption. 

 The quarter q-C in FIG. \ref{fig:Progression} has 300 nm of GaAs grown on a substrate that underwent Ga-assisted oxide desorption. As shown in FIG. \ref{fig:Progression} and in Table \ref{Tab:Growths}, q-C has a greatly improved surface morphology, with smaller RMS roughness compared to q-B. However, as seen in the BBS image, that morphology was not uniform across the entire quarter wafer piece. The smoothest area measured by AFM was located in the thin band region bordering the outside edge of the wafer.

\subsubsection{Mounting with a Sapphire Diffuser}

The nonuniform surface morphology observed in q-C (FIG. \ref{fig:Progression}) is attributed to a temperature gradient across the wafer during annealing and growth. Such gradients are a common challenge when using quarter wafers in substrate holders originally designed for full wafers. A system with similarly mounted substrate pieces was reported to have temperature variations of up to 15 °C across the wafer piece \cite{jackson_thermal_2007}. For the wafer q-C, the smoothest region was confined to a narrow band, with rougher areas on either side, indicating that precise control of growth temperature is critical for achieving a uniform and smooth surface morphology. This highlights the sensitivity of GaAs growth on (111)B substrates to temperature variations.

\begin{figure}
\centering
\includegraphics[width=\columnwidth]{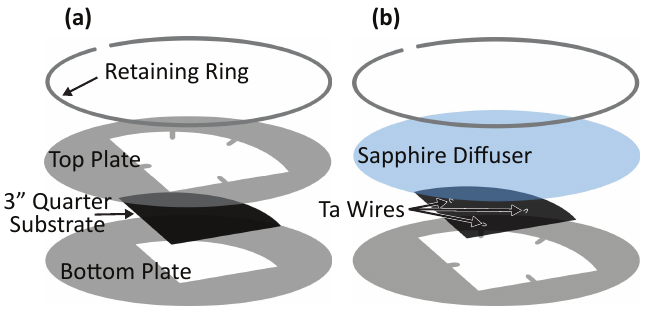}
\caption{Schematics showing two methods for securing 3" quarter wafers with Mo plates for mounting in full 3" substrate holders. The entire assembly secured by a Mo retaining ring, with the growth surface facing downwards. (a) The standard method with two Mo plates where the top plate has four tabs providing additional support. (b) The alternate method where the Mo plate with four tabs is now on the bottom and the top plate has been replaced with a sapphire diffuser. Three Ta wires minimize thermal contact between the sapphire and the quarter wafer. Figure sourced from Ref. [\onlinecite{shi_molecular_2021}].}
\label{fig:Mounting}
\end{figure}

To accommodate quarter wafer pieces in standard 3" wafer holders, the pieces are typically sandwiched between two circular molybdenum (Mo) plates with cutouts that expose both the growth surface and the backside, as shown in FIG. \ref{fig:Mounting} (a). The growth surface is facing downwards in FIG. \ref{fig:Mounting}. The cutouts are designed to minimize contact with the wafer while providing sufficient support. However, the manual alignment of the wafer and Mo plates introduces variability, as improper positioning or warping of the plates can result in uneven contact across the wafer. This significantly affects the uniformity of heating during growth.

Additionally, the Mo plates partially cover the wafer edges on the growth side. Reflections of thermal radiation from the Mo plates can result in multiple passes of radiation through the wafer, further complicating the thermal profile. These factors make the system highly complex and sensitive, with run-to-run variations in temperature gradients that are often difficult to attribute to specific causes.

To address this issue in subsequent growths, the top Mo plate was replaced with a 3" sapphire optical diffuser wafer, as shown in FIG. \ref{fig:Mounting} (b). The bottom Mo plate was substituted with the top Mo plate because it only contacts the wafer piece on the four supporting tabs. Short (\textless 0.5 cm) and thin tantalum (Ta) wires were placed between the sapphire substrate and the quarter wafer. These wires minimize thermal contact between the wafer and the sapphire substrate. In this configuration, the wafer has limited direct contact with the holder. The entire assembly was secured with a Mo retaining ring.

This mounting adjustment was effective in improving the surface morphology uniformity, as demonstrated by q-D in FIG. \ref{fig:Progression}. 


\subsubsection{\label{subsec:Surfactant} Surfactant-Mediated Growth}

Another technique sometimes used to improve surface morphology in challenging epitaxial growths is the surfactant-mediated epitaxy. A surfactant is a substance introduced to the surface during growth that is not incorporated in the epitaxial layer. It must remain mobile on the surface and readily segregate to stay at the growth front with minimal incorporation into the deposited layer, as noted in Ref. [\onlinecite[p.~45]{ida_sadeghi_realization_2021}]. The surfactant can enhance surface morphology by altering surface energy and influencing adatom mobility.

Using indium (In) as a surfactant is convenient because the In effusion cell is already available in our system. Although the literature on In as a surfactant is limited, it has been shown to be effective in nitride growths, where it enhances Ga adatom mobility\cite{qwah_indium_2022}. For our growths, In is expected to be an appropriate surfactant because our growth temperatures are well above 550°C, where In re-evaporation becomes significant \cite{grey_growth_1995}. 

For the q-E quarter, In flux and Ga flux were both supplied in comparable amounts during the 300 nm GaAs deposition. The In flux was set for a growth rate of $R_{InAs}= 1.15$ Å/s (without accounting for re-evaporation).  The BBS image of q-E in FIG. \ref{fig:Progression} demonstrates that such an excessive In flux can be detrimental to surface morphology. The RMS roughness was 2.50 nm, significantly higher than for q-D.

\begin{figure}
\centering
\includegraphics[width=\columnwidth]{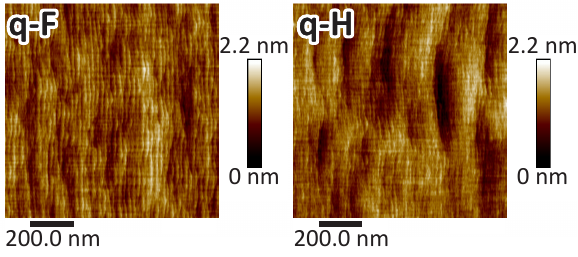}
\caption{1 µm$^2$ AFM scans from the smoothest regions of the quarter wafers q-F and q-H. Both wafers underwent Ga-assisted oxide desorption and subsequent layer growth with low flux In surfactant. The quarter q-F has 300 nm of GaAs grown, yielding an R$_{RMS}$ = 0.256 nm, while q-H has both a 200nm GaAs buffer layer and an additional 300 nm Al$_{0.55}$Ga$_{0.45}$As layer, yielding an R$_{RMS}$ = 0.302 nm.}
\label{fig:OxideDesorptionGrowthAFM}
\end{figure}

For the subsequent growth of 300 nm of GaAs on quarter q-F, the In flux was reduced to 0.20 Å/s growth rate while the Ga flux was still set for 1.0 Å/s. The resulting epitaxial layer exhibited a uniform morphology with distinct ordered atomic steps. FIG. \ref{fig:OxideDesorptionGrowthAFM} shows the AFM scan from the center of q-F, with an RMS roughness of 0.256 nm. This smooth morphology was present across the entire wafer piece.


Next, we focused on optimizing conditions for growing Al$_{0.55}$Ga$_{0.45}$As. For quarters q-G, q-H and q-I, the goal was to grow the GaAs Buffer Layer and Al$_{0.55}$Ga$_{0.45}$As Etch Stop 1 layer from Table \ref{tab:Schematic}. An additional capping layer of 5 nm of GaAs was then deposited on top to prevent rapid oxidation of the Al$_{0.55}$Ga$_{0.45}$As. The Al$_{0.55}$Ga$_{0.45}$As growth rate was always nominally $R_{Al_{1-x}Ga_xAs} = 2.23$ Å/s. It turned out that the optimal growth conditions used for GaAs were not as well optimized for Al$_{0.55}$Ga$_{0.45}$As.

For q-G, Al$_{0.55}$Ga$_{0.45}$As was grown using the same In surfactant flux as for the optimized GaAs layer in q-F. This growth yielded an RMS roughness of 0.548 nm in a 1 µm$^2$ scan, which is more than twice that for q-F. Furthermore, on a larger scale, the surface was covered with numerous pits up to 2.5 nm in depth.

To examine the effect of substrate temperature, quarter q-H was grown under identical conditions to q-G, with exception of the Al$_{0.55}$Ga$_{0.45}$As layer for which the temperature was increased to 630 °C. At this higher growth temperature, the surface morphology across the quarter q-H, shown in FIG. \ref{fig:Progression}, is very nonuniform. Several regions scatter light significantly, as seen in the corresponding BBS photo. These regions correspond to the four tabs used to support the quarter wafer in the substrate holder. The temperature gradient across the wafer has become significant at this higher growth temperature, leading to rough surface morphology around the supporting tabs. After examination with AFM, the smoothest region was identified near the bottom left (from the orientation in FIG. \ref{fig:Progression}), yielding an RMS roughness of 0.302 nm over a 1 µm$^2$ scan shown in FIG. \ref{fig:OxideDesorptionGrowthAFM}. This nearly matches the roughness of the GaAs layer from q-F, but the uniformity remains a significant issue, with most of the wafer showing pits similar to those observed for q-G.

To isolate the effects of the In surfactant during Al$_{0.55}$Ga$_{0.45}$As growth, q-I was grown at 590 °C, under the same conditions as q-G, but without In flux during the Al$_{0.55}$Ga$_{0.45}$As layer deposition. Despite this difference, q-G and q-I showed comparable surface morphologies, with RMS roughness values of 0.548 nm and 0.599 nm, respectively. These results suggest that, at the growth temperature of 590 °C, under the selected growth conditions, the In surfactant had no significant impact on the surface morphology of the Al$_{0.55}$Ga$_{0.45}$As layer.

To confirm the material compositions, several points on the quarter wafers were scanned with XRD. The XRD scans for q-F were used to investigate how much In was incorporated into the grown GaAs layer. Since the growth was done without rotation, the composition of In$_x$Ga$_{1-x}$As varied between $x=0.017$ and $x=0.025$ across the quarter wafer, indicating that some In incorporated into the material.

For q-I, the XRD scans were used to determine the composition of the Al$_y$Ga$_{1-y}$As layer, which was grown when no In flux was applied. Across the wafer, the Al composition varied between $y=0.680$ and $y=0.571$, with the target value at the center of the wafer being $y=0.55$, showing that the Al flux was higher than intended.

For q-G, the XRD scans are more difficult to simulate because there are three group III elements in the In$_x$Al$_y$Ga$_{1-x-y}$As layer. To simplify the analysis, it was assumed that the Al composition profile across q-G would be approximately the same as for the quarter q-I since only a small amount of In was expected to incorporate into the material, and the growth conditions were otherwise the same. The XRD scans for q-G can then be used to estimate the In composition in the In$_x$Al$_y$Ga$_{1-x-y}$As layer. With such assumptions, the amount of In that incorporated into the In$_x$Al$_y$Ga$_{1-x-y}$As layer appears to vary across the wafer between $x=0.023$ and $x=0.033$.

From the XRD scans of q-G and q-I, we see that a greater amount of In incorporated into the In$_x$Al$_y$Ga$_{1-x-y}$As than into the In$_x$Ga$_{1-x}$As layers for the same growth temperature and In flux. It is important to note that the nominal growth rate of Al$_{0.55}$Ga$_{0.45}$As was more than twice that of GaAs. So, In incorporation rate into Al$_{0.55}$Ga$_{0.45}$As is much higher than for GaAs for the same growth condition. 


\section{Conclusion}

We reported on progress in the optimization of GaAs and Al$_{0.55}$Ga$_{0.45}$As on (111)B substrates, leading to improvements in surface morphology. These improvements included a decrease in RMS roughness and a notable enhancement in the uniformity of surface morphology across most of the wafer pieces. In the pursuit of atomic flatness, it was discovered that the traditional thermal oxide desorption method typically used for GaAs substrates significantly roughened the (111)B substrate surface, similar to the effect seen for GaAs (001) substrates. However, unlike for GaAs (001), a smooth surface could not be recovered with subsequent GaAs deposition because of the sensitivity of growth modes on the (111)B plane to surface orientation

To address this, a procedure was developed using Ga pulses, in the absence of As flux, to reactively remove the oxides present on the substrate wafers without significant degredation of the surface quality.

Highly sensitive in-situ characterization techniques were used to monitor and control this oxide desorption process. Diffuse laser scatter (DLS) measurement was demonstrated to be a powerful technique for monitoring the evolution of surface morphology in situ, with a very high sensitivity for atomic-scale roughness. Additionally, the RHEED signal provided complementary real-time observations of oxide removal. 

Another key observation was that using a sapphire wafer as an optical diffuser when mounting quarter wafer pieces reduced the temperature gradients, significantly improving the uniformity of the surface morphology for GaAs layers. 

Finally, the use of In as a surfactant further improved the surface morphology of GaAs (111) growths. When combined with a higher growth temperature, it also enhanced the surface morphology of Al$_{0.55}$Ga$_{0.45}$As, achieving local atomic-scale smoothness and promoting the desired step-flow growth mode. However, uniformity of the surface morphology at this elevated temperature remains an issue. The sensitivity of the (111) growth modes to substrate temperature was much stronger in the case of quarter q-H where Al$_{0.55}$Ga$_{0.45}$As layer with In surfactant was grown at a temperature of 630 °C. In this case, even the optical diffuser did not prevent significant variation in substrate morphology across the wafer. We expect that owing to much smaller temperature variations, the epitaxial growth of AlGaAs layers on full 3" substrates would result in a large area of atomically smooth surface under otherwise similar growth conditions. 

The level of surface roughness achieved in both GaAs and Al$_{0.55}$Ga$_{0.45}$As layers on the GaAs (111)B substrates is already acceptable for the fabrication of QOMs or similar devices. This research paves the way for the fabrication of a variety of devices based on GaAs/AlGaAs heterostructures with (111)B crystal orientation. The fabrication of QOMs from these optimized growths is underway and preliminary results are very promising.



\begin{acknowledgments}

The University of Waterloo's QNFCF facility was used for this work. This infrastructure would not be possible without the significant contributions of CFREF-TQT, CFI, ISED, the Ontario Ministry of Research \& Innovation and Mike \& Ophelia Lazaridis. Their support is gratefully acknowledged.

In addition to the above institutions, this work was directly supported by the Natural Sciences and Engineering Research Council of Canada (NSERC) and funding from the Canada First Research Excellence Fund (CFREF).
\end{acknowledgments}

\section*{Data Availability Statement}
The data that support the findings of this study are available from the corresponding author upon reasonable request.

\nocite{*}
\section*{\label{sec:Declaration}Author Declarations}
The authors have no conflicts to disclose.

\section*{\label{sec:References}References}
\bibliography{Ref}

\end{document}